# Magnetic properties and phase diagrams of mixed spin-1 and spin-1/2 Ising model on a checkerboard square structure: A Monte Carlo study

Maen Gharaibeh[1,*], Mohammad H.A. Badarneh[1,2], Samah Alqaiem[1], Abdalla Obeidat[1] and Mohammad-Khair Qaseer[1]

[1]Department of Physics, Jordan University of Science and Technology, Irbid 22110, Jordan

[2]Science Institute, University of Iceland, 107 Reykjavík, Iceland

*Corresponding author: magh@just.edu.jo

**Abstract**

The magnetic properties and phase diagrams of the mixed spin-*1* and spin-*1/2* Ising model on a checkerboard square structure have been studied using the Monte Carlo simulations based on the Metropolis update protocol. The system consists of four quartets of alternative spin configuration. The effect of the exchange interactions *J* and crystal field *D* on the magnetic properties, critical and compensation temperatures, susceptibility, and specific heat of the system have been investigated. The phase diagrams, *T-J* and *T-D,* for different values of the exchange interactions and crystal field have been examined. We found that the compensation temperature starts to evolve for $J_1 < 0.6$ of the spin-1 assembly and for $J_2 > 1.6$ for spin-1/2. On the other hand, the effect of ferrimagnetic coupling does not show any threshold value; in this case, the compensation temperature is constant. Regarding the crystal field strength, the threshold value of $D > -0.8$ has been observed. We obtain the *N-*, *Q-* and *P*-type compensation behaviors in the system. We have observed that the phase diagrams exhibit only a second-order phase transition to a paramagnetic phase; hence, the system does not show the tricritical point. The magnetic hysteresis cycles of the mixed spin-*1* and spin-*1/2* on a checkerboard square structure for different values of the exchange interactions, temperatures, and crystal fields have been found. Finally, the system exhibits the superparamagnetic behavior for a fixed value of the temperature and crystal field.

**1. Introduction**

The mixed-spin Ising models have attracted a lot of interest over the last few decades in the magnetism community due to their huge industry and technology applications such as motors and magnetic recording [1-4]. These models are used to study the critical behaviors [5], magnetic properties [6], and to obtain the compensation temperature [7, 8]. The existence of compensation has been verified in different systems with various spins and different structures [9-12]. Therefore, this field is very active in both in solid-state and statistical physics. Even though the Ising model is considered one of the simplest models that deal with interactions, models with different spins magnitude are useful and give a deep understanding of ferromagnetic materials' magnetic behavior, which are a key ingredient in many applications in magnetic recording. It has been shown by many scientists that different two-dimensional mixed spin Ising models have a wide variety of unusual physical properties [13, 14].

After the pioneering work of Onsager in solving the two dimensional Ising model analytically [15], many attempts to get an exact solution on different decorated systems such as triangular [16], honeycomb [17], kagome [18], bathroom-tile [19], and ruby [20] lattices have been obtained. Other lattices such as the checkerboard pattern is not an easy task to accomplish analytically. For this reason, we applied the Monte Carlo simulations to study the magnetic properties on a checkerboard square pattern taking into account the effect of anisotropy interaction and the interaction of spin with an external magnetic field. Therefore, different structures with different exchange coupling have been proposed to give richness in this type of research [21-25].

The Monte Carlo simulations are used to study different structures of mixed spin atoms such as spin-2 and spin-3/2 Ising model on a diamond-like decorated square [26]. The ground-state phase diagrams, the effect of the reduced transition temperature and the crystal field, the magnetic hysteresis cycles of the mixed spin-5/2 and spin-2 Ising model on a decorated square lattice have been studied using the Monte Carlo simulations [27]. The effective field theory has been used to study the critical phenomena in a mixed spin-1 and spin-2 Ising model on a honeycomb structure [17] and the dynamic phase transition properties for the mixed spin-(1/2, 1) Ising model on a square lattice [28] as well as two nanoscaled thin films with dilution at the surfaces [29].

This work was motivated by the research done by Girovsky, Jan et al. [30]. They have produced a wafer-thin ferrimagnet, in which hydrocarbon compounds with different magnetic centers, composed of manganese and iron, arrange themselves to form a checkerboard pattern after applying them on a gold surface. They have also proven that the manganese and iron magnetism is of different strengths and appears in opposing directions, which is a characteristic feature of a ferrimagnet. In the present work, we will investigate the effect of the exchange interaction coupling and the crystal field on the thermal magnetization, critical and compensation behavior, susceptibility, the specific heat, hysteresis loops, and phase diagrams of a ferrimagnetic checkerboard square structure composed of spin-*1* and spin-*1/2* within the framework of Monte Carlo simulations. The structure we adopted here has ferrimagnetic coupling through the interface of the checkerboard domains. This paper is organized as follows: in section 2, we define our model and give the related formulation. The results and discussions have been given in detail in section 3. Finally, section 4 is devoted to our conclusions.

## 2. Model and formalism

*2.1 Lattice structure and Hamiltonian*

We consider an Ising ferrimagnetic checkerboard square lattice (see Fig. 1) of size $L \times L$ with $L$ is the crystal side length. The checkerboard lattice has four quarters with equal side lengths. Each quarter is composed of *S*-type of spins-*1* or *σ*-type of spins-*1/2*. The sublattice-*A* of the higher spin is represented in the first and the third quarters (blue filled circles in Fig. 1), and the sublattice-*B* consists the lower spin is represented in the second and the fourth quarters (red filled circles in figure 1). The coordination number of each site is 4.

The Hamiltonian terms consist of *S-S*, *σ-σ* ferromagnetic interactions and *S-σ* ferrimagnetic interaction, spin - external filed interaction, and crystal field interaction with *S*-type spins. The Hamiltonian is given by:

$$H = -J_1 \sum_{\langle i,j \rangle} S_i S_j - J_2 \sum_{\langle l,k \rangle} \sigma_l \sigma_k - J_{12} \sum_{i,l} S_i \sigma_l - B \sum_i (S_i + \sigma_i) - D \sum_i S_i^2 \qquad (1)$$

Where $J_1$, $J_2$, and $J_{12}$ stand for the coupling constant between the spins *S-S*, *σ-σ*, and *S-σ*, respectively. The summation indices $\langle i,j \rangle$ and $\langle l,k \rangle$ denote the summations over all nearest-neighbor spins *S-S* and *σ-σ*, respectively. The fourth and the last terms in the Hamiltonian are the external magnetic field acting on the *S*-type and *σ*-type spins and the crystal field interaction acting on the *S*-type spins, respectively. In our simulations, we are using the usual Ising variables *S=±1,0* and *σ=±1/2*. In this paper, we will take $J_1 > 0$ and $J_2 > 0$ to ensure a ferromagnetic interaction between the same type of spins, while we take $J_{12} < 0$ to ensure an antiferromagnetic interaction between the spins of different types. It is worth mention that the only ferrimagnetic interaction exists at the interfaces of the four sublattices. Therefore, we ought to use a large value of $J_{12}$ to ensure that neither *S*-type nor *σ*-type spins behave separately.

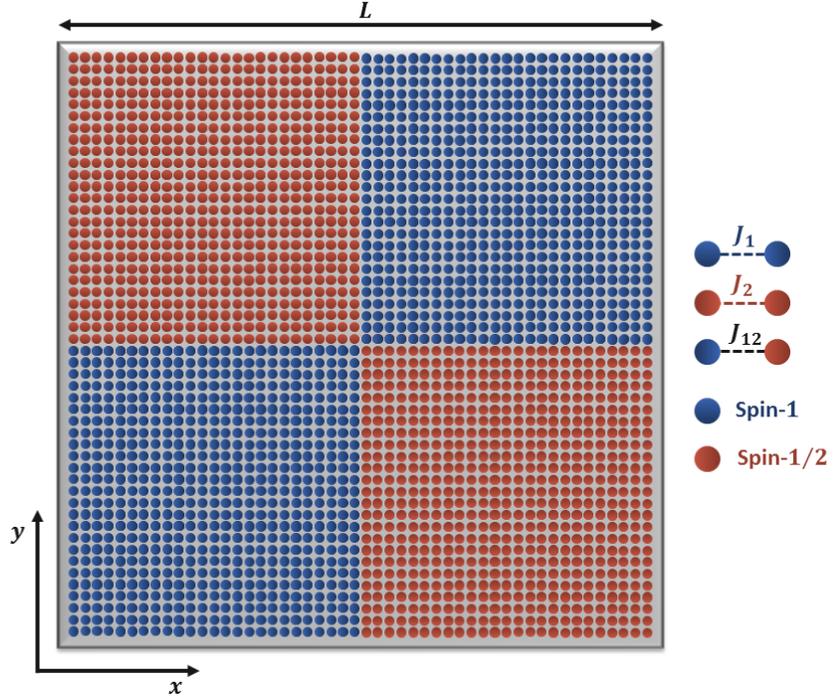

**Fig. 1.** Schematic representation of the checkerboard structure formed by sublattices *A* and *B* with spin-*1* and spin-*1/2*, respectively. The blue filled circles represent magnetic atoms (spin $S=\pm1,0$) in sublattice *A*, while the red filled circles represent magnetic atoms (spin $\sigma=\pm1/2$) in the sublattice *B*. The lines connecting the red and blue circles denote the nearest-neighbor exchange couplings ($J_1$, $J_2$, and $J_{12}$). The side length of the square structure is denoted by (*L*).

*2.2 Monte Carlo Simulation and Calculations*

To simulate the checkerboard ferrimagnetic system with great success, we implement the Monte Carlo simulation technique based on the metropolis algorithm [31]. We apply periodic boundary conditions in the *x* and y directions. To generate new configurations, we choose a spin at random state and then flip it. By making spin-flip attempts, each flip is accepted or rejected according to the metropolis algorithm. Data were generated using 1000000 Monte Carlo simulations to equilibrate the system, followed by 850000 Monte Carlo steps for each spin configuration. The results are reported for systems size *L*=50. Therefore, the total number of the spins in our simulation is $N_{tot}$=2500, which contains $N_A$=1250 spins of the *S*-type in the sublattice-*A* and $N_B$=1250 spins of the $\sigma$-type in the sublattice-*B*. We performed additional simulations with *L*=80 and *L*=100, but no significant differences were found from the results presented here. Calculation of the error is based on the method of blocks; the *L*-size is divided into $n_b$ blocks of length $L_b = L/n_b$. The number of blocks is chosen such that $L_b$ is higher than the correlation length. Therefore, error bars are calculated by grouping all the blocks, then taking the standard deviation [32]. In this paper, we did not use any reduced parameters. Hence, the temperatures of the system are measured in units of energy. Our program calculates the following parameters, namely:

The magnetization per site of the sublattice A and B can be calculated by

$$M_A = \frac{1}{N_A} \langle \sum_{i=1}^{N_A} S_i \rangle \qquad (2)$$

$$M_B = \frac{1}{N_B} \langle \sum_{l=1}^{N_B} \sigma_l \rangle \qquad (3)$$

$$M_{tot} = \frac{M_A + M_B}{2} \qquad (4)$$

The magnetic susceptibilities for each sublattice are given by:

$$\chi_A = N_A \beta (\langle M_A^2 \rangle - \langle M_A \rangle^2) \qquad (5)$$

$$\chi_B = N_B \beta (\langle M_B^2 \rangle - \langle M_B \rangle^2) \qquad (6)$$

and the total susceptibility is

$$\chi_{tot} = \frac{\chi_A + \chi_B}{2} \qquad (7)$$

Where $\beta = 1/k_B T$, $T$ is the absolute temperature, and $k_B$ is the Boltzmann factor. For simplicity, we set $k_B = 1$.

Finally, we have calculated the specific heat of the system as follows:

$$\frac{C}{k_B} = \frac{\beta^2}{N_{tot}} (\langle H^2 \rangle - \langle H \rangle^2) \qquad (8)$$

At the compensation temperature, $T_{comp}$, the sublattice magnetizations cancel each other, and the system's total magnetization is zero. Hence, to determine $T_{comp}$ from the computed magnetization data, a crossing point of the absolute value of the sublattice *A* and *B* magnetizations needs to be determined under the following condition:

$$|M_A(T_{comp})| = |M_B(T_{comp})| \tag{9}$$

$$\text{sign}\left(M_A(T_{comp})\right) = -\text{sign}\left(M_B(T_{comp})\right) \tag{10}$$

with $T_{comp} < T_c$, where $T_c$ is the critical temperature. Equations (9) and (10) indicate that the absolute values of the sublattice magnetizations are equal to each other; however, sign of them is different at the compensation point, $T_{comp}$. In this paper, the second-order phase transition is determined from the maxima of the susceptibility and specific heat curves.

## 3. Results and discussions

This section will present the results of the magnetic and thermodynamic properties of the mixed-spin ferrimagnetic checkerboard structure obtained using the Monte Carlo simulations. We have observed the influence of Hamiltonian parameters on the phase diagrams, magnetization, susceptibility, and specific heat of the system in the absence of the external magnetic field and finally obtained hysteresis loops.

### 3.1. Phase diagrams

To explore the effect of the exchange coupling $J_1$ on the critical and compensation temperatures, $T_c$ and $T_{comp}$, respectively, we have plotted in Fig. 2 the phase diagram of the system for different exchange coupling $J_1$ in the absence of the external magnetic field. Fig. 2 shows the phase diagram ($T$, $J_1$) for $J_2 = 2.0$, $J_{12} = -3.0$, $D = 0$ and $B = 0$. From this figure, we can see that, as $J_1$ increases, the critical temperature $T_c$ remains constant for $J_1 <$ 0.4 and increases linearly for $J_1 > 0.6$. The compensation temperature $T_{comp}$ when it exists also increases linearly to a threshold value of $J_1$ (for instance, the threshold value is $J_1 = 0.6 \pm 0.0247$). Exceeding this threshold value, the compensation temperature disappears. Hence, the threshold value of $J_1$ determines whether the system can exhibit the compensation behavior or not. It is worth noting that, as $J_1$ increases, the variations in the compensation temperature has a rapid increase compared to the critical temperature. This is due to the fact that increasing the value of $J_1$ leads to a fast ordering of the sublattice-*A*. A threshold point exists at which these two temperatures $T_c$ and $T_{comp}$ start to merge. This threshold point occurs at $J_1=0.6\pm0.0247$. We can remark that the same behavior has been observed in Ref. [33-35].

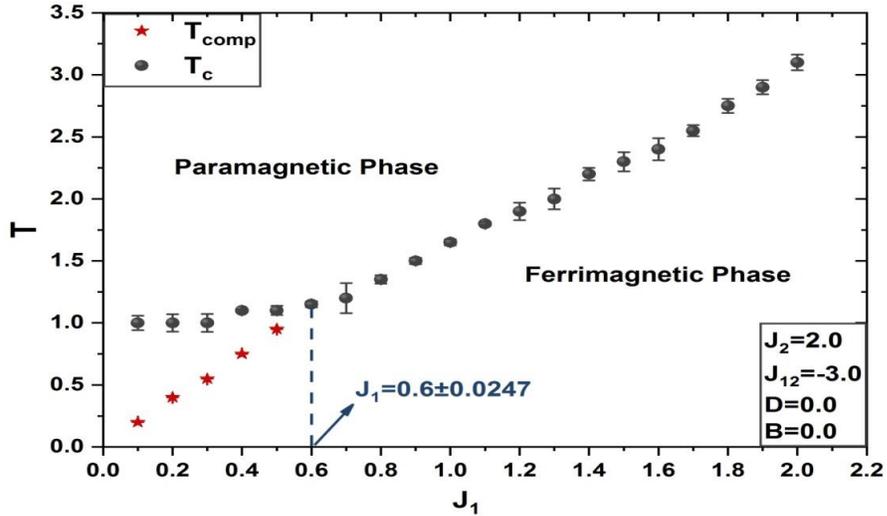

**Fig. 2.** The phase diagram of the system in $(T, J_1)$ plane for $J_2 = 2.0$, $J_{12} = -3.0$, $D = 0$, and $B = 0$. The dashed line intersects the horizontal axis at the threshold value of $J_1$. The error bars are smaller than the point markers.

Fig. 3 illustrates the influence of the exchange coupling $J_2$ on the critical and compensation temperatures, $T_c$, and $T_{comp}$, respectively for $J_1 = 0.5$, $J_{12} = -3.0$, $D = 0$, and $B = 0$. We can see that there exists a threshold value of $J_2$ (for instance, the threshold value is $J_2 = 1.6 \pm 0.01966$), which determines whether the system can exhibit the compensation behavior or not. Below the threshold value, the compensation temperature is no longer exists, and the critical temperature slightly increases. At the threshold value of $J_2$, a bifurcation of zero magnetization curves occurs. That is, as $J_2$ increases above the threshold value, the compensation temperature appears and remains unchanged, whereas the critical temperature has a gradual increase with increasing $J_2$. This result can be explained as follows: we recall that the stronger exchange coupling $J_2$ tends to make the spins of the sublattice-B ordered at higher temperatures, which increases the value of the critical temperature of the system at which a second-order phase transition occurs. Comparing figures 2 and 3, in the $T$-$J_1$ phase, the $T_{comp}$ exists for temperatures below the threshold value, while in the $T$-$J_2$ phase, the $T_{comp}$ exists for $T$ greater than the threshold value. Moreover, in the $T$-$J_2$ phase, no variations for $T_{comp}$ with $J_2$.

Also, since the value of $J_1$ is fixed, increasing the value of $J_2$ does not change the crossing point of the absolute value of the sublattice $A$ and $B$ magnetizations. However, it only changes the general behavior of the magnetization tail of the sublattice-$A$. Hence, the compensation temperature remains constant. Similar behavior has also been observed in nanoparticles with hexagonal core-shell structure [36] and a mixed Ising ferrimagnet with spins (3/2, 5/2) alternating on a square lattice [37].

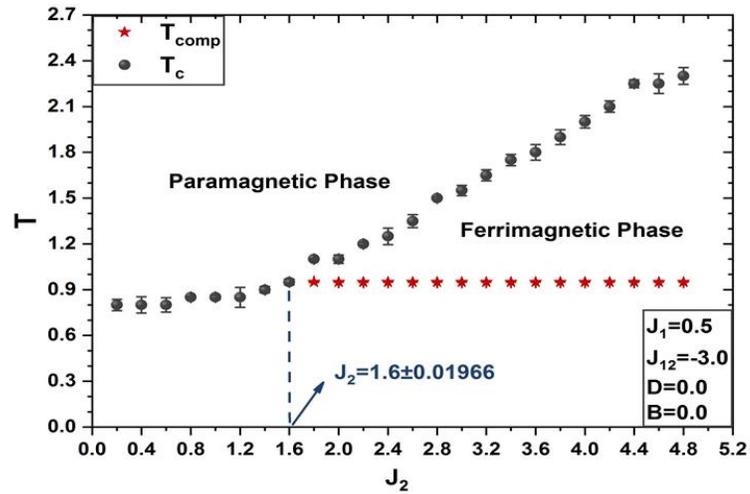

**Fig. 3.** The phase diagram of the system in ($T$, $J_2$) plane for $J_1 = 0.5$, $J_{12} = -3.0$, $D = 0$, and $B = 0$. The dashed lines intersect the horizontal axis at the threshold value of $J_2$. The error bars are smaller than the point markers.

To explore the effect of exchange ferrimagnetic interaction $J_{12}$, we plot in Fig. 4 the phase diagram of the system in the ($T$, $J_{12}$) plane for $J_1 = 0.2$, $J_2 = 2.0$, $D = 0$, and $B = 0$ at different values of $J_{12}$. We can see that the exchange interaction $J_{12}$ does not affect the compensation temperature, while the critical temperature remains almost constant. Here, we take the value of $J_1$ in the compensation regime of the $T$-$J_1$ phase, and $J_2$ is the compensation regime of the $T$-$J_2$ phase. The similar phenomena are observed in mixed spin (1-1/2-1) three layers system of cubic structure [34].

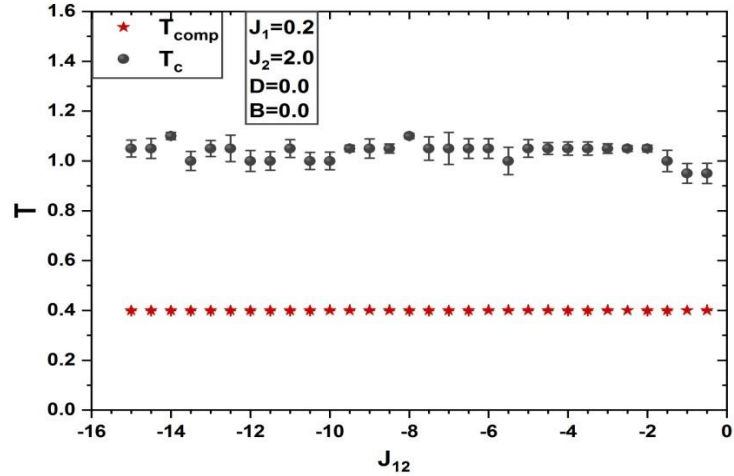

**Fig. 4.** The phase diagram of the system in ($T$, $J_{12}$) plane for $J_1 = 0.2$, $J_2 = 2.0$, $D = 0$, and $B = 0$

To study the effect of the crystal field $D$, Fig. 5 shows the phase diagram ($T$, $D$) for $J_1 = 0.5$, $J_2 = 3.0$, $J_{12} = -3.0$ and $B = 0$. We can see that for $D = 0$, the system exhibits critical and compensation temperatures, which can also be confirmed from Fig. 4. When the crystal field for $D < 0$ decreases, the compensation temperature decreases until it disappears below a threshold value $D = -0.8 \pm 0.0003$ at which, below this threshold value, no compensation exists, while increasing $D$ for positive values of the crystal field results in increasing the compensation temperature of the system. The figure shows a gradual increase in $T_{comp}$ with increasing $D$ and reaches a constant value for $D > 3$. We also observe that the critical temperature of the system slightly changes with varying the crystal field. Hence, we can conclude that the crystal field plays a major role in the appearance of the system's compensation behavior, which is beneficial for the thermomagnetic recording materials [37, 38]. Therefore, it is not just the magnetic interaction that influenced the $T_{comp}$ and $T_c$, but the crystal field $D$ has a significant impact on both temperatures. In our earlier work on different structures, we ignored the crystal field because of its weak significance [14, 34]. The same behavior has also been observed in nanoparticles with hexagonal core-shell structure [36] as well as a Blume Capel ferrimagnetic core/shell nanoparticle with spherical shape [39].

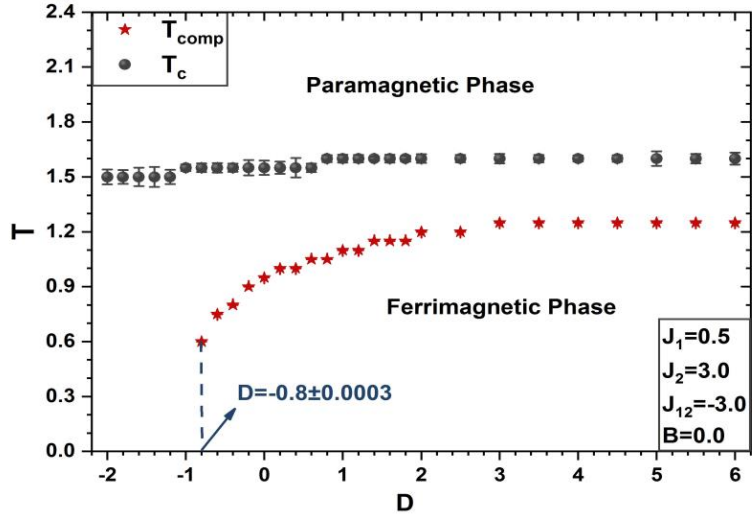

**Fig. 5.** The phase diagram of the system in *(T, D)* plane for $J_1$=0.5, $J_2$=3.0, $J_{12}$=-3.0, and *B*=0. The dashed lines intersect the horizontal axis at the threshold value of *D*. The error bars are smaller than the point markers.

*3.2. Magnetic properties*

In order to confirm the results obtained in the previous figures of phase diagrams (Fig. 2, Fig. 3, Fig. 4, and Fig. 5), we will present in this section the general trend of the behavior of the magnetization, susceptibility as well as the specific heat of the system as a function of temperature for selected values of the system parameters. The values of $J_1$ and $J_2$ have been chosen to be within the range of existence of $T_{comp}$ of the *T-$J_1$* and the *T-$J_2$* phases discussed above.

In Fig. 6, we show the sublattice magnetizations $M_A$, $M_B$, total magnetization $M_{tot}$ of the system, total susceptibility, and specific heat as a function of temperature. We have chosen $J_1 = 0.5$, $J_2 = 3.0$, $J_{12} = -3.0$ and B = 0 for different positive values of the crystal field $D = 0, 0.5, 2.0$ and $5.0$. In Fig. 6a, we can note that the magnetization curves of the sublattice A and B decrease monotonically, from the saturation values $|M_A| = 1$ and $|M_B| = 0.5$ at zero temperature, as the temperature increases, and diminishes to zero at the critical temperature ($T_c=1.6$), which results in a second-order phase transition to a paramagnetic phase. We also note that $M_B$ changes more slowly than that of $M_A$, which is due to the fact that the stronger exchange coupling $J_2$ tends to make the spins of the sublattice-B ordered at higher temperatures. In Fig. 6b, we plot the system's total magnetization using Eq.4 as a function of temperature. The magnetization curves show two magnetizations zero points. The first point at lower $T$ corresponds to the compensation temperature at which the system's total magnetization reduces from the saturation ($|M_{tot}| = 0.25$) to zero.

The second one at higher $T$ corresponds to the critical temperature at which the second-order phase transition to a paramagnetic phase occurs. In both figures, the crystal field's effect shifted the magnetization curves to the right toward higher temperatures, suggesting that the compensation temperature increases with increasing the crystal field's positive value. In contrast, the second magnetization zero points of all magnetization curves stay fixed, which indicates that the system's critical temperature remains constant. To understand the crystal field's effect on the compensation temperature, we recall that the crystal field interacts with the spins in the sublattice-$A$. Using large positive values of the crystal field suggests that a higher temperature is needed to disorder the sublattice-$A$ by forming magnetic domains inside it. Hence, the crossing point of the absolute value of the sublattice A and B magnetizations increases, thereby increasing the system's compensation temperature. Similar behavior of $M_A$, $M_B$ and $M_{tot}$ as a function of temperature has been confirmed in a ferrimagnetic mixed-spin (2, 5/2) Ising system on a layered honeycomb lattice [40], two dimensional mixed-spin (1, 1/2) graphene-like Ising nanoparticle [41] and ferrimagnetic mixed-spin (2, 5/2) system on a bipartite square lattice [42]. The variation of the total susceptibility $\chi_{tot}$ as a function of temperature is depicted in Fig. 6c. One can remark that, for each value of the crystal field, two peaks of susceptibility occur. The first peak occurs due to the abrupt drop of $M_A$, as shown in Fig. 6a, and its location coincides with the compensation temperature location. The second peak occurs at the critical temperature at which phase transition to the paramagnetic phase occurs, which indicates the second-order phase for magnetic materials. With the increase of the positive value of the crystal field, the location of the first peak moves right, which also proves that the compensation temperature increases as the value of the crystal field increases. The location of the second peak remains fixed, which indicates that the critical temperature is the same as the value of the crystal field increases. The double-peak phenomena appear in the susceptibility curves have well confirmed within the framework of the mean-field theory based on Bogoliubov inequality for the Gibbs free energy [43]. Furthermore, Monte Carlo simulations have also well confirmed the double-peak phenomena in the susceptibility curves [27, 34, 35, 42, 44]. The influence of the crystal field's positive values on the specific heat $C$ is shown in Fig. 6d. We observed two peaks for each value of the crystal field. The first peak at lower $T$ occurs at the compensation temperature, while the second peak at higher $T$ occurs at the transition temperature. These results are in agreement with the previous results, Fig. 6 (b and c). We can remark that a similar double-peak phenomenon in $C$ curves have been observed in a ferrimagnetic mixed-spin (1, 3/2) Ising nanowire with hexagonal core-shell structure [23], ferrimagnetic mixed-spin (5/2, 2) on a bipartite square lattice [42], mixed spin-5/2 and spin-3/2 Ising model on a square lattice [45], triple layer spin (1-1/2-1) cubic system [34], ferrimagnetic mixed-spin (3/2, 5/2) in a graphene layer [46].

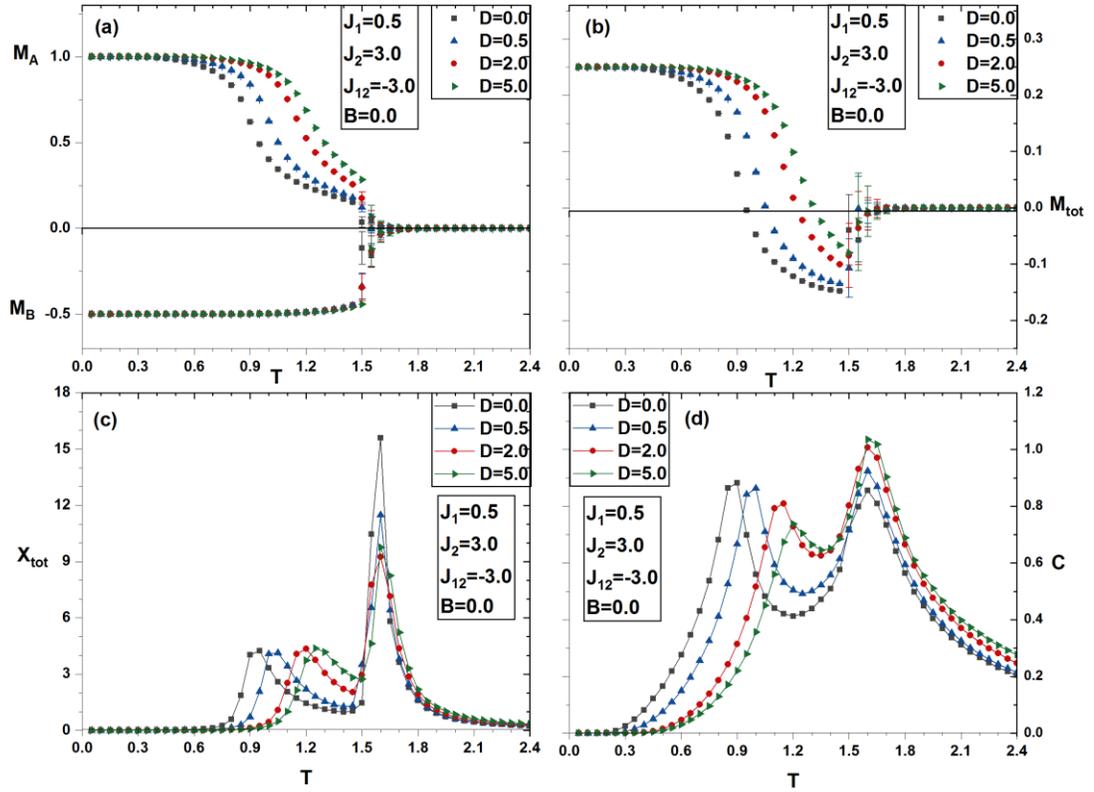

**Fig. 6.** The temperature dependencies of (a) sublattice magnetizations, (b) total magnetization, (c) total susceptibilities, (d) specific heat, for $J_1 = 0.5$, $J_2 = 3.0$, $J_{12} = -3.0$, $B = 0$ and different positive values of $D$ ($D = 0$, 0.5, 2.0 and 5.0).

The influence of the crystal field's negative values on the sublattice magnetizations $M_A$, $M_B$, total magnetization $M_{tot}$ of the system, total susceptibility, and specific heat as a function of temperature is depicted in Fig. 7. We have chosen $J_1 = 0.5$, $J_2 = 3.0$, $J_{12} = -3.0$ and $B = 0$ but with negative values of the crystal field $D = 0$, -0.4, -0.8 and -1.0. Fig. 7a shows that the sublattice A and B's magnetization curves decrease monotonically, from the saturation values $|M_A| = 1$ and $|M_B| = 0.5$ at zero temperature as the temperature increases and diminishes to zero at the critical temperature ($T_c = 1.6$). The figure also shows that $M_A$ changes more rapidly than $M_B$ because of the weaker exchange coupling $J_1$. More magnetic domains appear in the sublattice-$A$ as the temperature increases, results in the fast disordering of the sublattice-$A$. In Fig. 7b, we plot the system's total magnetization using Eq.4 as a function of temperature. Two magnetizations zero points in the magnetization curves were observed. The first one denotes the compensation temperature, whereas the second one corresponds to the critical temperature. It is worth mentioning that the first magnetization zero points of each magnetization curve move left, which suggests that the compensation temperature decreases with increasing the negative value of the crystal field.

In contrast, the second magnetization zero points of all magnetization curves are fixed, suggesting that the system's critical temperature remains constant. By using large negative values of the crystal field, more magnetic domains inside the sublattice-$A$ appear even at low temperatures, which means a very low temperature is required to ensure a perfectly ordered sublattice-$A$, hence, the crossing point of the absolute value of the sublattice $A$ and $B$ magnetizations decreases, thereby the compensation temperature of the system decreases. In Fig. 7c, we plot the total susceptibility $\chi_{tot}$ using Eq.7 as a function of temperature. We note that for each value of the crystal field, two peaks of susceptibility occur. The first peak occurs due to an abrupt drop of $M_A$, as shown in Fig. 7a, and its location coincides with the location of the compensation temperature. The second peak occurs at the critical temperature. With the increase of the crystal field's negative value, the location of the first peak moves to the left, which also proves that the compensation temperature decreases as the negative value of the crystal field increases.

In contrast, the second peak location remains fixed, which indicates that the critical temperature is the same as the negative value of the crystal field increases. Finally, to confirm the results presented in Fig. 7 (b and c), we plot the specific heat $C$ of the system using Eq.8 as a function of temperature, as shown in Fig. 7d. Two peaks were observed in the specific heat curves for each value of the crystal field, and the location of the first (second) peak coincides with the location of $T_{comp}$ ($T_c$).

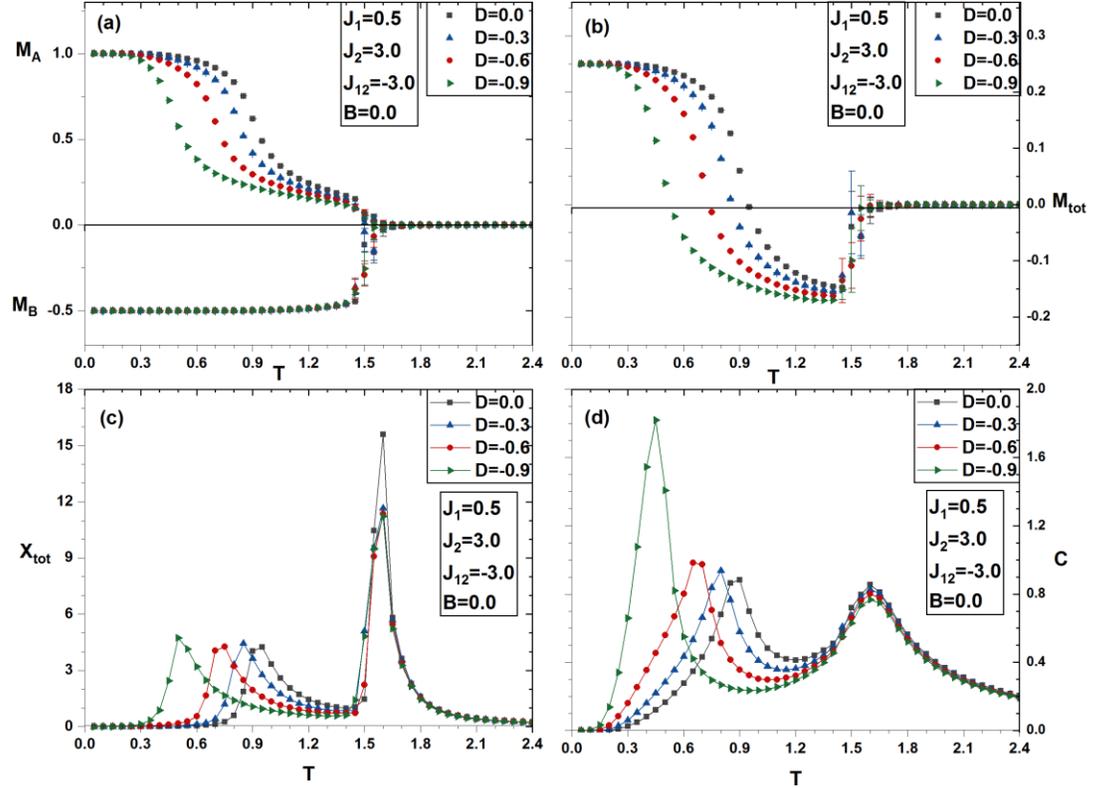

**Fig. 7.** The temperature dependencies of (a) sublattice magnetizations, (b) total magnetization, (c) total susceptibilities, (d) specific heat, for $J_1 = 0.5$, $J_2 = 3.0$, $J_{12} = -3.0$, $B = 0$ and different negative values of $D$ ($D = 0.0$, -0.3, -0.6 and -0.9).

Fig. 8 shows the system's total magnetization as a function of temperature for selected values of the system parameters in the absence of the external magnetic field. In Fig. 8(a) we varied $J_1$ for $J_2 = 2.0$, $J_{12} = -3.0$, $D = 0$ and $B = 0$. The magnetization curves present N-, Q- and P-type compensation behaviors in the system for $J_1 \leq 0.4$, $J_1 = 0.6$ and $J_1 = 0.9$, respectively as classified in the Néel theory [47]. Besides, one can observe that for $J_1 < 0.6$, the system exhibits critical and compensation temperatures, while for $J_1 \geq 0.6$, the system exhibits only critical temperature. For a very low-temperature $T$, both sublattices

*A* and *B* are fully ordered, thus leading to the saturation value of the total magnetization of the system $M_{tot} = \frac{1+(-1/2)}{2} = 0.25$ by Eq (4). The negative sign in the previous equation appears because spins in both sublattices point in an opposite direction due to the effect of the ferrimagnetic coupling. We present in Fig. 8(b) the effect of $J_2$ on the general trend of the behavior of the system's total magnetization for $J_1 = 0.5$, $J_{12} = -3.0$, $B = 0$, and $D = 0$. In this figure, we can note that below the critical temperature, all the magnetization curves have the same zero magnetization point, which represents the compensation temperature. Hence, the compensation temperature does not change as the value of $J_2$ increases, while the second zero magnetization point of all curves move right, which proves that the critical temperature increases as $J_2$ increases. Also, the figure shows *N*-type magnetization, where the $J_1$ and $J_2$ are within the $T_{comp}$ range for the phases mentioned above of T-$J_1$ and T-$J_2$. To investigate the influence of the crystal field *D* on the system's total magnetization, we have shown this effect in Fig. 8 (c) by varying *D* for $J_1 = 0.5$, $J_2 = 3.0$, $J_{12} = -3.0$, and $B = 0$. The figure shows *N*-type magnetization as described in the Néel theory [47], where $J_1$ and $J_2$ are in the $T_{comp}$ range. One can notice that for large positive values of the crystal field *D*, the compensation temperature increases. Using large positive values of the crystal field makes the spin-flip in sublattice-*A* harder, leads to a slower change in $M_A$ with the increase of temperature. Hence, the crossing point of the absolute value of the sublattice *A* and *B* magnetizations increases to higher values, therefore, increasing the system's compensation temperature. It is also worth noting that the compensation temperature decreases for large negative values of the crystal field. The decrease is attributed to the fact that using large negative values of the crystal field makes the spin-flip in sublattice-*A* easier, which leads to a faster change of $M_A$ with the increase of temperature. Hence, the crossing point of the absolute value of the sublattice *A* and *B* magnetizations decreases, thereby decreasing the system's compensation temperature. Consequently, the behavior of the magnetization curves shown in Fig. 8 has well confirmed by means of Monte Carlo simulations in a mixed Ising ferrimagnet with spins (3/2, 5/2) alternating on a square lattice [37], ferrimagnetic mixed-spin (2, 5/2) Ising system on a layered honeycomb lattice [40], ferrimagnetic nanocube with a spin-3/2 core surrounded by a spin-1 shell layer [48], the mixed spin-7/2 and spin-3 ferrimagnetic Ising system on a square lattice [49], nano-dicoronylene like-structure in a Blume-Capel model [50], ferrimagnetic mixed spin-2 and spin-5/2 Ising system on a honeycomb lattice [51].

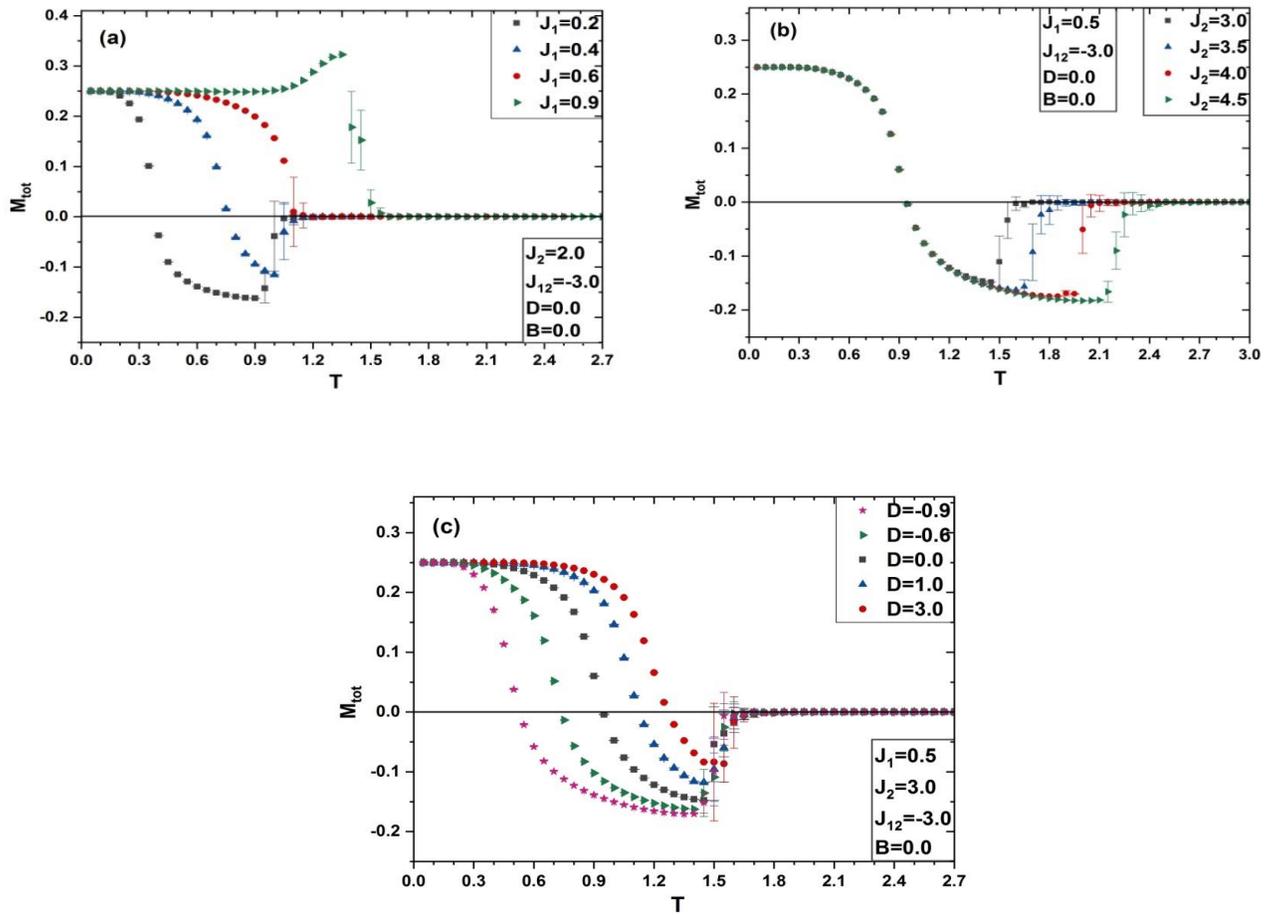

**Fig. 8.** The temperature dependencies of the total magnetization of the system for (a) $J_2 = 2.0$, $J_{12} = -3.0$, $B = 0$ and $D = 0$ with $J_1 = 0.2, 0.4, 0.6$ and $0.9$, (b) $J_1 = 0.5$, $J_{12} = -3.0$, $B = 0$ and $D = 0$ with $J_2 = 3.0, 3.5, 4.0$ and $4.5$, (c) $J_1 = 0.5$, $J_2 = 3.0$, $J_{12} = -3.0$, $B = 0$ with $D = -0.9, -0.6, 0, 1.0$ and $3.0$

*3.3. Hysteresis loops*

In this part, we investigated the effect of $J_1$, $J_2$, $J_{12}$, and $D$ on the magnetic hysteresis cycles for a ferrimagnetic checkerboard structure. Our results are shown in Figures. 9 to 13.

Fig. 9 shows the effect of temperature on the magnetic hysteresis cycles. We have plotted in Fig. 9 (a) the total magnetization (left panel), (b) sublattices magnetization (right panel) versus the external magnetic field for $J_1 = 1.0$, $J_2 = 2.0$, $J_{12} = -3.0$, and $D = 0$ at $T = 0.8, 1.0$ and $1.8$. We can observe that the hysteresis cycle's surface loop decreases with increasing the temperature. If the temperature increases, the magnetic domains will start to appear in the system, so less work is needed, hence the area decreases, namely, the loop disappears. We observe that the system enters into the superparamagnetic phase at $T = 1.8$. This behavior is observed in the mixed spin-3/2 and spin-2 Ising model on a diamond-like decorated square [26], the mixed spin-7/2 and spin-3/2 Ising model located in alternating sites of a square lattice [52], a mixed Spin-3/2 and Spin-1/2 Ising Ferrimagnetic System [11], mixed spin-1/2 and spin-3/2 Ising nanoparticles system within the framework of the EFT with correlations [53].

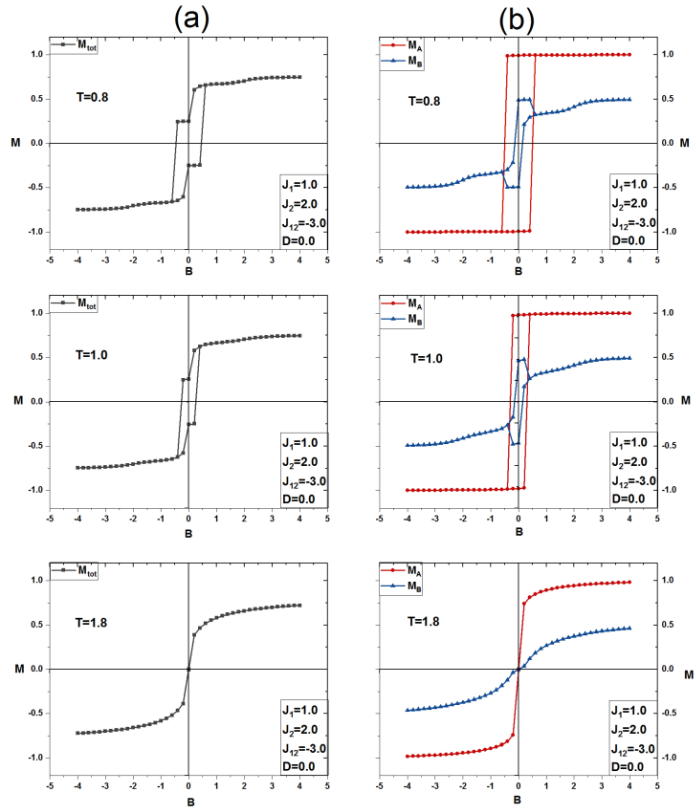

**Fig. 9.** The magnetic hysteresis loops with different temperature $T$ ($T = 0.8$, 1.0, and 1.8) for (a) total magnetization, (b) sublattice $A$ and $B$ magnetization when $J_1 = 1.0$, $J_2 = 2.0$, $J_{12} = -3.0$ and $D = 0$

In order to study the effect of $J_1$ on the magnetic hysteresis cycles, we have plotted in Fig. 10 (a) the total magnetization (left panel), (b) sublattice magnetization (right panel) versus the external magnetic field for $J_2 = 2.0$, $J_{12} = -3.0$, $D = 0$ at $T = 0.8$ with $J_1 = 1.5$, 2.0 and 2.5. From the figure, one can see that as $J_1$ increases, the area of the loop also increases; hence, the coactivity of the system is significantly improved. Also, two steps hysteresis exists. The obtained results are similar to those obtained in Ref. [53-55].

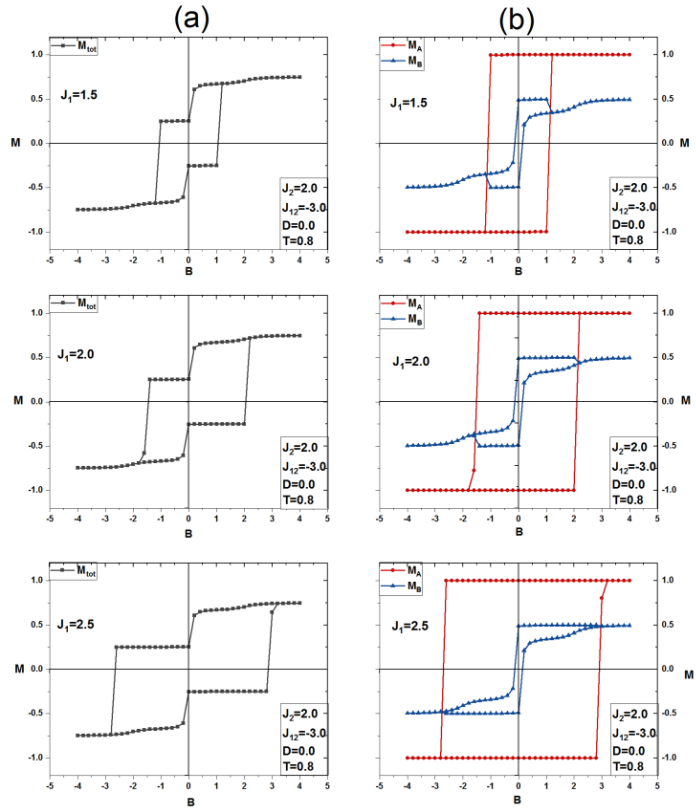

**Fig. 10.** The magnetic hysteresis loops with different exchange coupling $J_1$ ($J_1$ = 1.5, 2.0 and 2.5) for (a) total magnetization, (b) sublattice $A$ and $B$ magnetization when $J_2$ = 2.0, $J_2$ = -3.0, $D$ = 0 and $T$ = 0.8

Fig. (11, 12) shows the effect of $J_2$ on the magnetic hysteresis cycles. We have plotted in Fig. (11, 12) (a) the total magnetization (left panel) (b) sublattice magnetization (right panel) versus the external magnetic field for $J_1$ = 1.0, $J_{12}$ = -3.0, $D$ = 0, and $T$ = 0.8 with $J_2$ = 0.5, 1.0 and 5.0. We found that the shape of the hysteresis loop changes when increasing the value of $J_2$; the system has a single hysteresis loop for $J_2$ = 0.5 and 1.0, while at very high $J_2$ = 5.0 two steps hysteresis loop exist. In addition, similar multiple hysteresis loop behaviors have also been discovered in many theoretical studies of magnetic nanowires [56-58] nanoparticles [59, 60] and molecule magnets [61]. The interesting multiple hysteresis behaviors in experiments have also been discussed in nanoparticles [62].

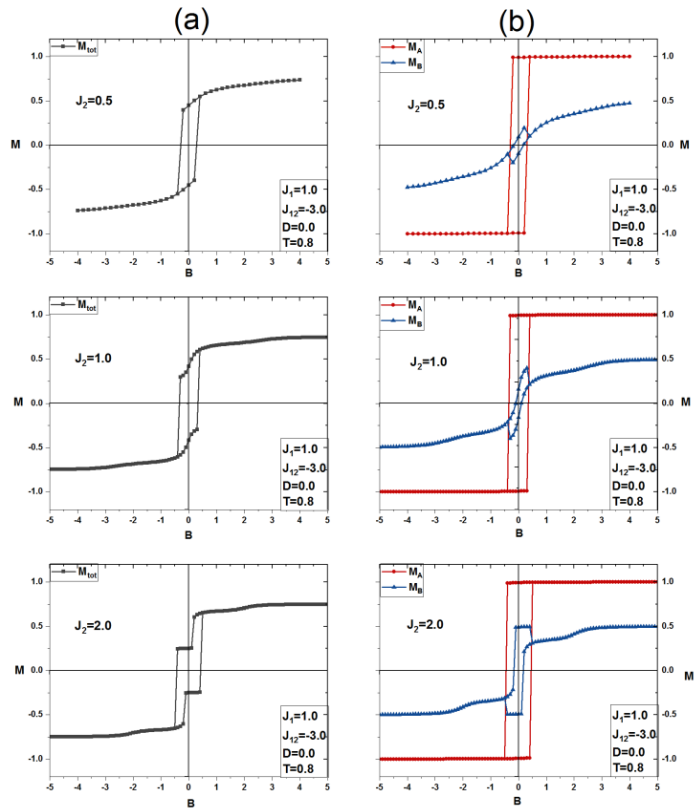

**Fig. 11.** The magnetic hysteresis loops with different exchange coupling $J_2$ ($J_2 = 0.5$, 1.0 and 2.0) for (a) total magnetization, (b) sublattice $A$ and $B$ magnetization when $J_1 = 1.0$, $J_{12} = -3.0$, $D = 0$ and $T = 0.8$

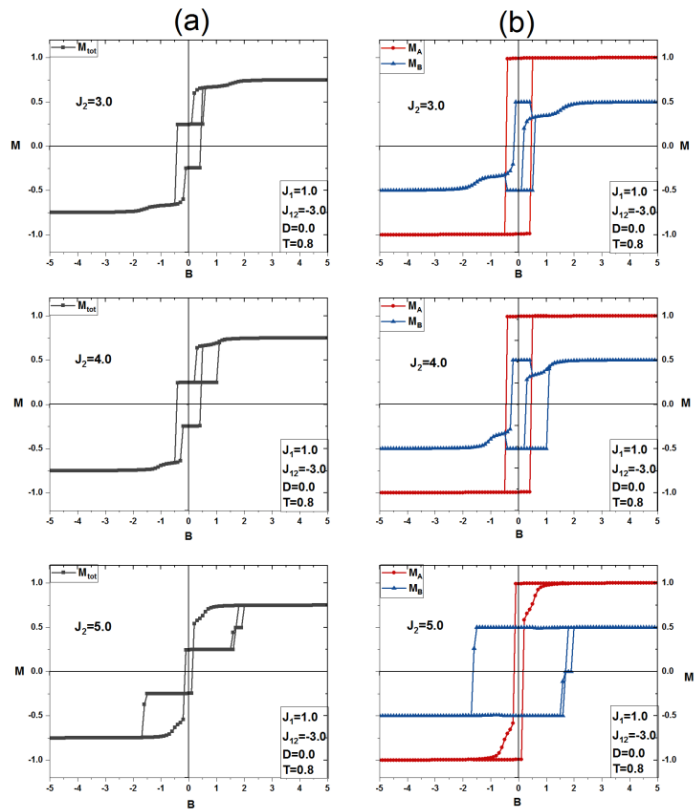

**Fig. 12.** The magnetic hysteresis loops with different exchange coupling $J_2$ ( $J_2$ = 3.0, 4.0 and 5.0) for (a) total magnetization, (b) sublattice $A$ and $B$ magnetization when $J_1$ = 1.0, $J_{12}$ = -3.0, $D$ = 0 and $T$ = 0.8

The effect of ferrimagnetic coupling does not affect the shape of the hysteresis loop; for this reason, we preferred not to show it in the context.

Fig. 13 shows the effect of $D$ on the magnetic hysteresis cycles, we have plotted in Fig. 13 (a) the total magnetization (left panel) (b) sublattice magnetization (right panel) versus the external magnetic field for $J_1 = 1.0$, $J_{12} = 2.0$, $J_{12} = -3.0$ and $T = 0.8$ with $D = -2.0$, $-1.0$ and $1.0$. The figure shows that the system exhibits a superparamagnetic behavior at $D = -2$, and as we are increasing $D$, the magnetization reveals two-step hysteresis with a wider width. Therefore, for such a checkerboard structure, this confirms the essential feature that $D$ plays a major role in the appearance of $T_{comp}$ (see Fig. 5). Similar behavior has been observed in Ref. [11, 53, 55, 63, 64].

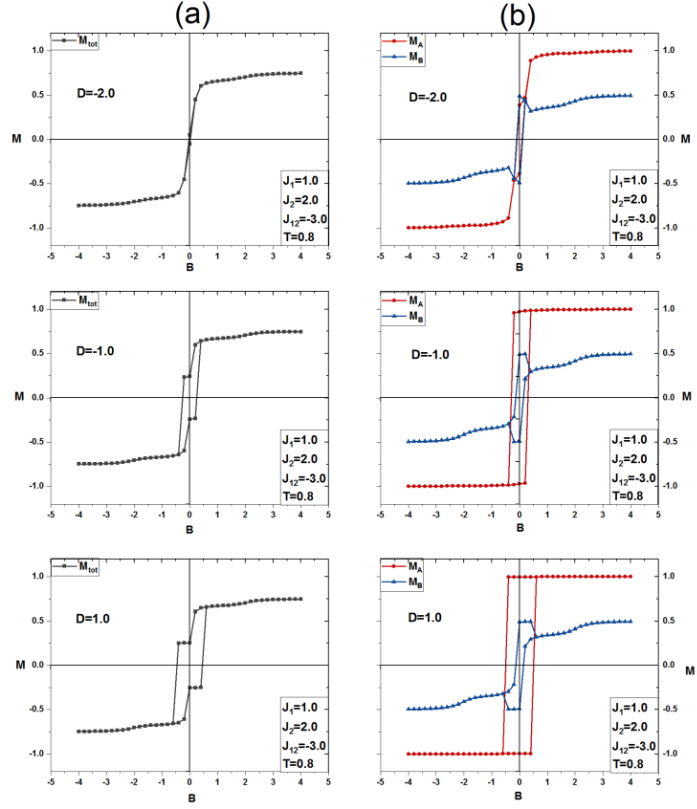

**Fig. 13.** The magnetic hysteresis loops with different $D$ ($D$ = -2.0, -1.0 and 1.0) for (a) total magnetization, (b) sublattice $A$ and $B$ magnetization when $J_1$ = 1.0, $J_2$ = 2.0, $J_{12}$ = -3.0 and $T$ = 0.8

Finally, to explore the effect of number of layers, we plot in Fig. 14 the phase diagram of the system in the ($T$, $L_z$) plane for $J_1$ = 0.5, $J_2$ = 3.0, $D$ = 0 and $B$ = 0. We have noticed that the compensation temperature reaches a saturation value for $L_z$ > 3, whereas the critical temperature reaches the saturation value for $L_z$ > 4. The value of both compensation and critical temperatures for $L_z$ = 1 can be compared with result in Fig. 3 in the text.

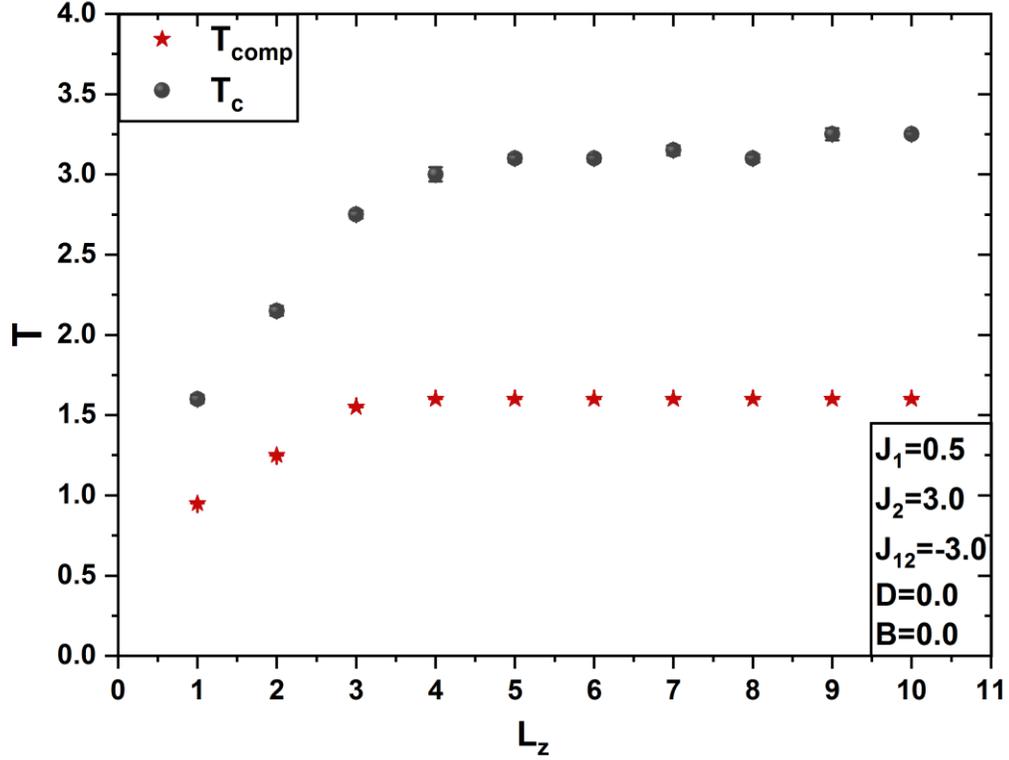

**Fig. 14.** The phase diagram of the system in (T, Lz) plane for $J_1 = 0.5$, $J_2 = 3.0$, $J_{12} = -3.0$, $D = 0$ and $B = 0$.

## 4. Conclusion

In the present work, we employed Monte Carlo simulations based on the Metropolis update protocol to study the effect of Hamiltonian parameters on the magnetic and thermodynamic properties and the phase diagrams of a ferrimagnetic checkerboard square structure. Our results show that the system exhibits a second-order phase transition to a paramagnetic phase. Threshold values of the exchange interactions ($J_1 < 0.6$ and $J_2 > 1.6$) and crystal field ($D > -0.8$) have been found in the phase diagrams, and these values are of great importance since they can determine whether the system can exhibit a compensation behavior or not. We have found that the ferrimagnetic coupling has no significant effect on the critical and compensation temperatures. We have observed that the system's critical temperature slightly changes with varying the crystal field, which in contrast to the compensation temperature. For high values of the crystal field ($D > 3$), the critical and compensation temperatures reach a saturation value and remain constant. The double-peak phenomenon has been observed in the susceptibilities and the specific heat curves, and the location of the first (second) peak coincides with the location of the compensation (critical) temperature. We have noticed that the compensation temperature is strongly linked with the Hamiltonian parameters of the system. The existence of the compensation point in our system makes it potential candidate for use in the area of thermo–magnetic data storage and magneto–optical recording media devices [37, 38, 65-67]. The critical temperature shows a gradual increase with increasing $J_1$ or $J_2$, and our simulations reveal the existence of $N$-, $Q$- and $P$-type compensation behaviors in the system. The coercive magnetic field increases with increasing the absolute value of exchange interactions $J_1$ and decreases with increasing the crystal field's negative values. The hysteresis loop pattern changes from a single loop into a double loop by increasing the value of $J_2$. The multiple hysteresis behaviors are of great interests in the applications of multi-state memory devices [52]. Also, the hysteresis loop area decreases as the system's temperature increases, whereas the ferrimagnetic coupling has no effects on the hysteresis loop area. Finally, the superparamagnetic behavior is observed at temperature $T = 1.8$ and crystal field $D = -2.0$.

## Acknowledgment

This work was supported by the Jordan University of Science and Technology under Grant (20200039).